# High CO tolerance of Pt/Ru nano-catalyst: insight from first principles calculation.

Sergey Stolbov[1], Marisol Alcántara Ortigoza[1], Radoslav Adzic[2]

Talat S. Rahman[1]

[1] *Department of Physics, University of Central Florida, Orlando, FL*

[2] *Chemistry Department, Brookhaven National Laboratory, Upton, NY*

*Abstract:* Density functional theory based calculations of the energetics of adsorption and diffusion of CO on Pt islets and on the Ru(0001) substrate show that CO has the lowest adsorption energy at the center of the islet, and its bonding increases as it moves to the edge of the island and further onto the substrate. Activation energy barriers for CO diffusion from the islet to the Ru surface are found to be lower than 0.3 eV making the process feasible and leading to the conclusion that this hydrogen oxidation catalyst is CO tolerant because of the spillover of CO from active Pt sites to the Ru substrate. We present the rationale for this effect using insights from detailed electronic structure calculations.

**Introduction**

Fuel cells are clean energy conversion devices which can be used in a wide variety of applications. However, there are significant obstacles on the way to their large scale implementations. Some of them can be overcome only based on understanding of microscopic mechanisms of interaction of atoms and molecules with solid surfaces and nano-structures. For example, Proton Exchange Membrane Fuel Cells (PEMFC's) operate with pre-reformed hydrogen which is usually obtained from hydrocarbons and therefore inevitably contains carbon monoxide. Even small traces of CO, remaining in the gas after purification, poison the commonly used Pt anode by blocking its active site which suppresses hydrogen oxidation. In Direct Methanol Fuel Cells (DMFC) the anode is used as a catalyst for both methanol reforming and for oxidation of the hydrogen so obtained. Although carbon monoxide released in the course of this reaction is supposed to be oxidized with OH obtained from admixed water, it still severely poisons the Pt anode. Both PEMFC and DMFC thus suffer from CO poisoning which is a major impediment in the efficient use of these fuel cells.

It is known that alloying of Pt with a second (and even third) metal element may reduce this poisoning effect. For instance, PtRu alloys are found to be more tolerant to CO than pure Pt [1]. Alloying Pt with Sn [2] and Mo [3] may improve the anode performance to some extent. However, these alloy anodes are also eventually affected by CO poisoning. Another disadvantage of these materials is high loading of expensive platinum. Not surprisingly the report that nanoclusters of Ru with sub-monolayer of Pt (PtRu$_{20}$) are much more tolerant to CO poisoning than commercial PtRu catalysts [4,5] has been welcomed with optimism. It is also important that the content of Pt in these novel materials is much lower than that in PtRu alloys. From estimate of the average diameter of Ru nanoparticles (2.5 nm) and Pt/Ru ratio, the authors conclude that the deposited Pt forms small islands (islets) on the facets of the Ru nanoparticles. The mechanism underlying the high CO tolerance of these particular nanostructures is yet to be understood, although attempts have been made to gain insight from related systems [6,7,]. For example, Koper et al. [6] have calculated from first principles CO adsorption energy $E_{ad}(CO)$ on clean Pt(111) and Ru(0001) surfaces, as well as on a Pt monolayer on Ru(0001) and a Ru monolayer on Pt(111). Since $E_{ad}(CO)$ is found to be the lowest for the case of a monolayer of Pt on Ru(0001) ($Pt_{ML}/Ru(0001)$), Koper et al. propose it to be the rationale for high CO tolerance for this system. First principles study of alloying effects on CO adsorption on Pt [7,!!!] suggest that strain induced by the second element modifies the electronic states of Pt in such a way that it causes a decrease in CO adsorption energy. Subsequent studies [5 – 9] have also upheld the view that enhanced Pt tolerance to CO poisoning is associated with a decrease in $E_{ad}(CO)$, as though reduction of $E_{ad}(CO)$ implies removal of CO from Pt active sites. Note that $E_{ad}$ for CO on these Pt surface alloys is still such that one expect CO to adsorb on the surface. In the case of $Pt_{ML}/Ru(0001)$, in which only Pt atoms are exposed to the surface, CO removal can be achieved only through enhancement of CO desorption. Desorption rate $R$ can be estimated using the transition state theory:

$$R = D_0 e^{-\frac{\Delta E}{kT}} \quad (1)$$

Where $D_0$ is the pre-factor, and, for this particular process, $\Delta E = E_{ad}(CO)$. Setting $D_0 = 10^{12}\ sec^{-1}$, which is a typical value for the pre-factor, $T = 350K$ (operation temperature for PEMFC), and taking $\Delta E = 1.11$ eV from Ref. 6, we obtain R ≈ $10^{-4}$ sec$^{-1}$. The desorption rate is thus quite low for $Pt_{ML}/Ru(0001)$, and it is expected to be even lower for other anodes, because of higher $E_{ad}(CO)$. Missing from this analysis is the consideration of CO diffusion rates which for anodes with inhomogeneous surfaces such as the $PtRu_{20}$ nanoparticles [4], may be the main factor contributing to the CO removal from Pt sites. Spillover of CO from Pt islands to the Ru substrate is mentioned in Ref. [8] as a possible mechanism for the high CO tolerance, but the argument contrives to be based on the assumption of weakened CO adsorption. However, weak adsorption does not necessarily guarantee high spillover. Conclusion about efficiency of the CO spillover needs to be based on activation energy barriers for CO diffusion in the system. One way of obtaining such information is through accurate first principles calculations of the system energetics based on density functional theory. In this work we have carried out such calculations of the energetics of adsorption and diffusion of CO on and off Pt nano-islands and the Ru(0001) substrate. Rationale for the obtained energetics is drawn from the calculated local densities of electronic states of the systems.

**Computational Details**

The first principles calculations reported in this paper have been carried out within the density functional theory (DFT) [10,11] using the plane wave pseudopotential method [12] as embodied in the code VASP [13] with ultrasoft pseudopotentials [14]. To maintain periodicity of the systems we used supercell comprising of a 5 layer Ru(0001) slab with a four, seven, or nine Pt-atom island, on one side, and vacuum layer of 15 Å. Most calculations also included the CO molecule adsorbed either on the island or on the Ru(0001) substrate. To diminish interaction between the periodic images of Pt islands, the supercell was extended along the (0001) surface making up the (4x4) superstructure. With such geometry the shortest distance between edges of neighboring islands equaled two Ru-Ru bond lengths. The supercell thus contained 80 Ru atoms, plus Pt atoms forming the island and a CO molecule. Brillouin zones were sampled with the (3x3x1) Monkhorst-Pack k-point meshes [15]. We used a kinetic energy cutoff of 400 eV for the wave functions and 700 eV for the charge density which provided sufficient computational accuracy for the oxygen containing structures. The Perdew-Wang generalized gradient approximation (GGA) [16] has been used for the exchange-correlation functional.

To achieve structural relaxation, a self-consistent electronic structure calculation was followed by calculation of the forces acting on each atom. Based on this information the atomic positions were optimized to obtain equilibrium geometric structures in which forces acting on atoms do not exceed 0.02 eV/Å.

**Results and Discussion**

We present here results for the 7-atom Pt islet on the Ru(0001) surface (7Pt/Ru(0001)). We have calculated the energies of CO adsorption on the top of the central (c-Pt) and edge (e-Pt) platinum atoms and on two non-equivalent (fcc and hcp) hollow sites, as well as on the Ru substrate site neighboring the Pt islet (n-Ru) and the next neighbor Ru site (nn-Ru). In addition, activation energy barriers have been calculated for CO diffusion from c-Pt to e-Pt, between two e-Pt (along the island edge), from e-Pt to n-Ru, and from n-Ru to nn-Ru sites. Fig. 1 shows the energetics calculated for the system with CO moving along the c-Pt – e-Pt – n-Ru – nn-Ru path.

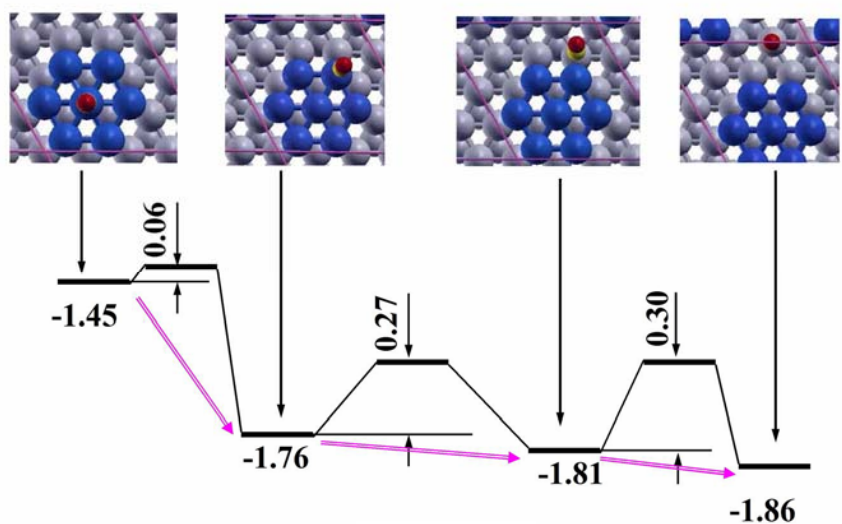

Fig. 1. Schematic illustration of energetics of CO diffusing from the center of Pt islet (c-Pt site) to its edge (e-Pt) and further to Ru substrate (n-Ru and nn-Ru sites). Red, yellow, blue and grey balls represent O, C, Pt and Ru atoms, respectively. Negative and positive numbers correspond to CO adsorption energies and CO diffusion energy barriers, respectively.

As illustrated in Fig. 1, CO bonding increases, as the molecule moves from the center of Pt island to its edge and further to the Ru(0001) substrate. This negative gradient of $E_{ad}(CO)$ along the c-Pt – e-Pt – n-Ru – nn-Ru path already indicates that CO molecule adsorbed on 7Pt/Ru(0001) would prefer leave the islet for the Ru substrate. Furthermore, we find $E_{ad}(CO)$ = -1.94 eV for CO on clean Ru(0001). This value is lower than those for CO adsorption on the n-Ru and nn-Ru sites suggesting that CO tends not only to leave the Pt islet, but also to move away from it.

The activation energy barrier for CO diffusion from c-Pt to e-Pt through the bridge site is found to be as low as 0.06 eV. The barriers for the rest of the considered path are also quite low. The highest n-Ru – nn-Ru barrier is 0.3 eV resulting in a diffusion rate $R \approx 5*10^7 \ sec^{-1}$ from Eq.1 with $D_0 = 10^{12} \ sec^{-1}$ and *T = 350K*. Clearly, this rate is much higher than that for CO desorption. We thus find the spillover of CO from Pt islets to Ru substrate to be a favorable process, which keeps active Pt sites available for hydrogen oxidation and hence provide the high CO tolerance of the $PtRu_{20}$ nanostructure. Interestingly, we find CO bonding to the Pt island atoms to be significantly stronger than its bonding to $Pt_{ML}$/Ru(0001) (we obtain $E_{ad}$(CO) = -1.15 for $Pt_{ML}$/Ru(0001), which differs slightly from -1.08 eV reported in Ref. 6). For the edge Pt atom it is even stronger than CO bonding to Pt(111) (-1.6 eV). Nevertheless, this system provides very efficient mechanism for CO removal from active Pt sites, and this mechanism originates not from weak CO bonding to Pt atoms, but from the negative adsorption energy gradient and low energy barriers for CO moving from the center of Pt island to its edge and further to the Ru(0001) substrate. The calculations have also been performed for a 4-atom Pt islet on Ru(0001), in which all Pt atoms belong to its edge. It is thus not surprising that the obtained $E_{ad}$(CO) = -1.78 eV is almost the same as for the edge Pt atoms in 7Pt/Ru(0001) (-1.76 eV). We can thus expect the same spillover scenario for smaller islands.

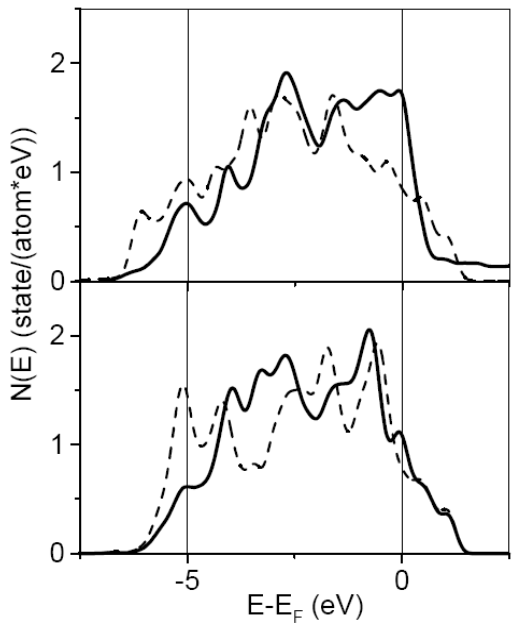

Fig. 2. Calculated Pt LDOS. Upper panel: Surface atoms of Pt(111) (solid line) and $Pt_{ML}$/(Ru(0001) (dashed line). Lower panel: e-Pt (solid line) and c-Pt (dashed line).

The rationale for the hierarchy in the CO energetics presented in Fig. 1, may be traced to the local densities of electronic states (LDOS) of 7Pt/Ru(0001). Indeed, as has been often discussed, hybridization between the metal surface atom and CO electronic states induces donation of electron density from the occupied bonding CO states to the metal atom and a back donation to the non-occupied anti-bonding CO states [17]. Such charge exchange results in CO – metal bonding whose strength depends on the energetic separation between metal *d*-band and the bonding and the anti-bonding CO states. The conclusion obtained here through explicit calculations are in quantitative agreement with those resulting from simple models [18,19] proposed a while back.

Our results are consistent with the above qualitative picture. The highest absolute values of $E_{ad}$(CO) are found for the n-Ru and nn-Ru sites. At the same time, these sites have high LDOS (> 1 state/(atom*eV)) from both sides of $E_F$ (see Fig. 2). Both the absolute value of $E_{ad}$(CO) and LDOS at $E_F$ (see Fig. 3) are much higher for Pt(111) than for $Pt_{ML}$/Ru(0001). Similarly, LDOS around $E_F$ and $E_{ad}$(CO) correlate for the e-Pt and c-Pt sites in 7Pt/Ru(0001). We can thus conclude that for the systems under consideration, LDOS around the Fermi-level controls the strength of the CO bonding to the surface.

Characteristics of Pt LDOS in 7Pt/Ru(0001) are mostly governed by the *d*-Pt – *d*-Pt and *d*-Pt – *d*-Ru hybridization. The Ru *d*-band is wider and has more non-occupied states than the Pt *d*-band. As a result, *d*-Pt – *d*-Ru hybridization redistributes Pt *d*-states into the region above $E_F$ and reduces the average Pt LDOS. It is clear that this effect is stronger if the Pt site has more Ru neighbors. Another factor affecting the LDOS of Pt is the total number of nearest neighbors: the larger this number, the wider is the *d*-band. Interplay of the above factors results in a variety of shapes and magnitudes of Pt LDOS shown in Fig. 3. In $Pt_{ML}$/Ru(0001) *d*-states of each Pt atom hybridize with *d*-states of 3 Ru atoms and has as many as nine nearest neighbors. This makes the Pt *d*-band wide, LDOS low, and CO bonding relatively weak. The e-Pt atoms in 7Pt/Ru(0001) have only six nearest neighbors which makes the $d$-band of e-Pt narrow and aligned to $E_F$. In addition, hybridization with three Ru neighbors increases the LDOS of e-Pt just above $E_F$. This causes strong CO bonding to the e-Pt sites. On the other hand, the c-Pt site has six Pt neighbors, just like Pt atoms in $Pt_{ML}/Ru(0001)$. However, in contrast to the latter, all the six are e-Pt atoms with the narrow $d$-band. As a result, the c-Pt site has both LDOS and $E_{ad}(CO)$ intermediate between e-Pt and $Pt_{ML}/Ru(0001)$.

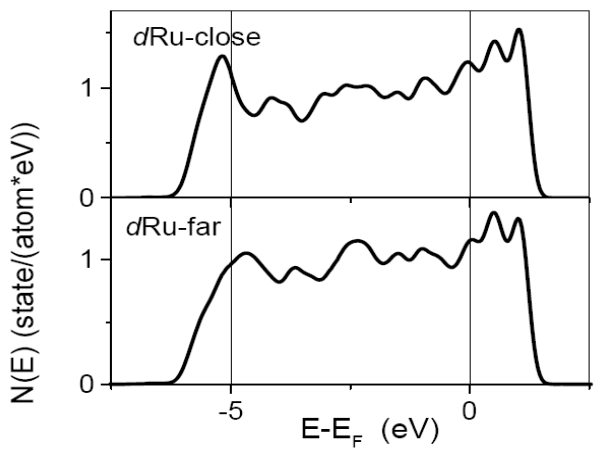

Fig. 3. LDOS calculated for n-Ru (upper panel) and nn-Ru (lower panel) sites.

For CO adsorbed on 7Pt/Ru(0001) we have also calculated LDOS of the C, O, c-Pt and e-Pt atoms. The three narrow, low-energy maxima, seen in Figs. 4 and 5, between -11 and -6 eV reflect strong $p$-C – $d$-Pt hybridization providing electronic charge donation from CO to Pt for CO adsorbed on both c-Pt and e-Pt. The main difference between plots shown in Figs. 4 and 5 is found in the energy region closer to $E_F$. A broad structure of C $p$-states in this region resulting from $pC$ – $dPt$ hybridization is found to be much more intensive for CO adsorbed on e-Pt than on c-Pt. Since these structures reflect the back donation of the electronic charge from Pt to CO, we conclude that as CO moves from c-Pt to e-Pt, the back donation is enhanced making CO bonding stronger.

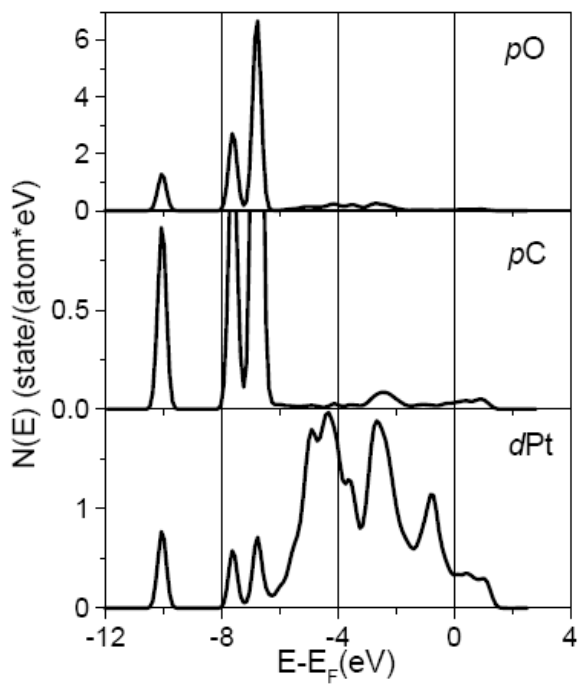

Fig. 4. LDOS calculated for CO adsorbed on the c-Pt site.

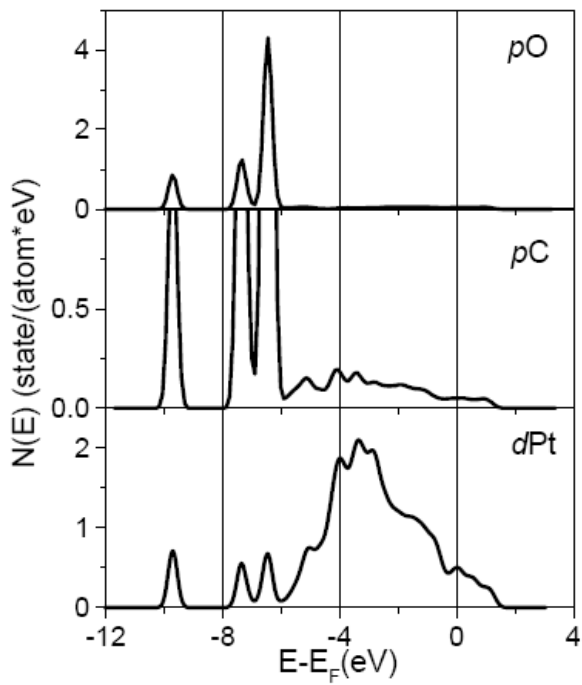

Fig. 5. LDOS calculated for CO adsorbed on the e-Pt site.

In summary, the first principles studies presented in this paper show that a fine balance among the number of nearest neighbors, ratio of Pt and Ru neighbors, and $d$-band widths of Pt and Ru results in a unique pattern of LDOS in 7Pt/Ru(0001), which causes gradual strengthening of CO bonding along the c-Pt – e-Pt – n-Ru – nn-Ru path. Such a gradient together with low activation barriers for CO diffusion across the system provide favorable condition for CO spillover from Pt islands to Ru substrate, which prevents poisoning of active catalytic Pt sites

**Acknowledgment:** Work supported in part by DOE under grant # DE-FG02-03ER15842.

**References:**

1. T. R. Ralph and M. P. Hogarth, Platinum Metals Rev., **46**, 117 (2002).
2. B. E. Hayden, M. E. Rendall, and O. South, JACS, **125**, 7738 (2003).
3. G. Samjeské, H. Wang, T. Löffler, and H. Baltruschat, Electrochim. Acta **47**, 3681 (2002).
4. S. R. Brankovic, J. X. Wang, and R. R. Adžic, Electrochem. Solid State Lett. **4**, A217 (2001).
5. S. R. Brankovic, J. X. Wang, Y. Zhu, R. Sabatini, J. McBreen, and R. R. Adžic, J. Electroanalytical Chem. **524-525,** 231 (2002).
6. M. T. Koper, T. E. Shubina, and R. A. van Santen, J. Phys. Chem. B **106**, 686 (2002).
7. M. Tsuda and H. Kasai, Phys. Rev. B **73**, 155405 (2006).
8. K. Sasaki, J. X. Wang, M. Balasubramanian, J. McBreen, F. Urbe, and R. R. Adžic, Electrochim. Acta **49**, 3873 (2004).
9. J. Greeley and M. Mavrikakis, Catalysis Today **111**, 52 (2006).
10. P. Hohenberg, W. Kohn, Phys. Rev. **136,** B864 (1964).
11. W. Kohn, L. J. Sham, Phys. Rev. **140**, A1133 (1965).
12. M. C. Payne, M. P. Teter, D. C. Allan, T. A. Arias, and J. D. Joannopoulos, Rev. Mod. Phys. **64**, 1045 (1992).
13. G. Kresse and J. Furthmuller, Phys. Rev. B **54**, 11 169 (1996)
14. D. Vanderbilt, Phys. Rev. B **41**, 7892 (1990).
15. H. J. Monkhorst and J. P. Pack, Phys. Rev. B **13**, 5188 (1976).
16. J. P. Perdew and Y. Wang, *Phys. Rev.* B **45**, 13244 (1992).
17. S.-S. Sung and R. Hoffmann, J. Am. Chem. Soc. **107,** 578 (1985).
18. G. Blyholder, J. Phys. Chem. **68**, 2772 (1964)
19. B. Hammer, Y. Morikawa, and J. K. Nørskov, CO chemisorption at metal surfaces and overlayers. Phys. Rev. Lett. **76**, 2141 (1996).